\documentstyle[aps,twocolumn,psfig,epsf]{revtex}

\input epsf

\tightenlines

\begin{document}

\title{\Large{Quark Number Susceptibility, Thermodynamic Sum Rule,
and Hard Thermal Loop Approximation}}

\author{Purnendu Chakraborty$^1$, Munshi G. Mustafa$^{1,2,}$, and
Markus H. Thoma$^3$}

\address{$^1$Theory Group, Saha Institute of Nuclear Physics, 1/AF Bidhan Nagar,
Kolkata 700 064, India}

\address{$^{2}$ Institut f\"ur Theoretische Physik, Universit\"at Giessen,
35392 Giessen, Germany}

\address{$^3$ Centre for Interdisciplinary Plasma Science, 
Max-Planck-Institut f\"ur extraterrestrische Physik,
P.O. Box 1312, 85741 Garching, Germany}

\maketitle

\begin{abstract}
The quark number susceptibility, associated with the
conserved quark number density, is closely related to the baryon and charge
fluctuations in the quark-gluon plasma, which might serve as
signature for the quark-gluon plasma formation in ultrarelativistic
heavy-ion collisions. Beside QCD lattice simulations, the quark number 
susceptibility has been calculated recently using a resummed perturbation 
theory (Hard Thermal Loop resummation). In the present work we show,
based on general arguments, 
that the computation of this quantity neglecting Hard Thermal Loop vertices
contradicts the Ward identity and 
violates the thermodynamic sum rule following from
the quark number conservation. 
We further show that the Hard Thermal Loop perturbation theory 
is consistent with
the thermodynamic sum rule.
\end{abstract}

\vspace{0.2in}

Dynamical properties of a many particle system can be investigated by 
employing an external probe, which disturbs the system only slightly 
in its equilibrium state, and by measuring the response of the system to
this external perturbation. A large number of experiments belong to this 
category~\cite{fluc} such as studies of various lineshapes, acoustic 
attenuation, and transport behavior. In
all these experiments, one probes the dynamical behavior of the
spontaneous fluctuations in the equilibrium state. In general, 
spontaneous fluctuations are related to correlation functions. 
They reflect the symmetries of the system,
which provide important inputs for quantitative  
calculations of complicated many-body systems~\cite{kubo}. 

Recently, screening and fluctuations of conserved quantities, such as baryon
number and electric charge, have been
proposed as a probe of the quark-gluon plasma (QGP) formation
in ultrarelativistic heavy-ion 
collisions~\cite{mcler,roland,asakawa,koch,hatsuda,rho,shuryak,zahed}. 
These fluctuations are closely related to
the quark number susceptibility (QNS)~\cite{mcler,hatsuda}, 
which measures the response of the number density
to an infinitesimal change of the quark chemical potential. 

The QNS has been considered within lattice gauge 
theory~\cite{gotlieb,gavai,gavai1}, in perturbative QCD
\cite{vuorinen}, as well as effective models of 
QCD~\cite{mcler,hatsuda,rho,shuryak,zahed}. Recently 
it has also been computed within the 2-loop approximately self-consistent
$\Phi-$derivable Hard Thermal Loop (HTL)
resummation in Ref.~\cite{rebhan1} and within one-loop HTL perturbation
theory (HTLpt) in Ref.~\cite{purn}. Whereas the result of Ref.~\cite{purn}
is in good agreement with lattice results, Ref.~\cite{rebhan1} overestimates 
the lattice data. Note, however, that the entropy and pressure obtained
within the 2-loop approximately self-consistent
$\Phi-$derivable method agree well with lattice data~\cite{blaizot1}.

In Ref.~\cite{rebhan} possible explanations for 
the different results of the various HTL approaches have been
discussed. As we will argue here, the neglect of the HTL vertices violates
the Ward identities and
the thermodynamic sum rule following from the conservation
of the quark number~\cite{fluc,mcler,hatsuda}. 
We show, based on general arguments, that the HTLpt, on the other hand, 
is consistent with the thermodynamic sum rule.

The HTL resummation technique allows a systematic gauge invariant
treatment of gauge theories at finite temperature and chemical potential
taking into account medium effects such as Debye screening, effective quark
masses, and Landau damping \cite{braaten}.
The generating functional which generates the HTL Green functions between
a quark pair and any number of gauge bosons can be written 
\cite{braaten,braaten0} as
\begin{equation}
\delta{\cal L} = m_q^2 {\bar \psi} \left \langle 
\frac{K \! \! \! \! /}{K\cdot D} \right \rangle \psi \ \ , \label{hg}
\end{equation}
where $K^\mu$ is a light like four-vector, $D_\mu$ is the covariant 
derivative, $m_q$ is thermal quark mass, and $\langle \ \ \rangle$ is the 
average
over all possible directions over loop momenta. This functional
is gauge symmetric and nonlocal and leads to the following
Dirac equation
\begin{equation}
D \! \! \! \! / \psi =  \Sigma \psi + \Gamma_\mu A^\mu \psi + A^\mu 
\Gamma_{\mu\nu}A^\nu \psi  
 +\cdots  \ \ \ , \label{hd}
\end{equation}
where we have suppressed the color index. 
In the HTL approximation the 2-point function, $\Sigma\sim gT$, 
(quark self-energy) is of
the same order as the tree level one, $S^{-1}_0(K) \sim K \! \! \! \! /
\sim gT$ (in the weak coupling limit $g\ll 1$), 
if the external momenta are soft, {\it i.e.} of the order $gT$. 
The 3-point function, 
$i.e.,$ the effective quark-gauge boson vertex, is given by
$g\Gamma_\mu=g(\gamma_\mu+\delta \Gamma_\mu)$, where $\delta \Gamma_\mu$ is the
HTL correction. 
The 4-point function, $g^2\Gamma_{\mu\nu}$, 
does not exist at the tree level and only appears 
within the HTL approximation \cite{braaten}.
These $N$-point functions, which are complicated functions of momenta and 
energies, are interrelated by Ward identities. This has important consequences
for the QNS (and other quantities) 
if one aims at a consistent perturbative expansion.

The HTLpt~\cite{andersen}
is an extension of the screened perturbation theory~\cite{karsch}  
in such a way that 
it amounts to a reorganization of the usual perturbation theory by 
adding and subtracting the nonlocal HTL term (\ref{hg}) in the 
action~\cite{braaten,taylor}. The added 
term together with the original QCD action is treated nonperturbatively 
as zeroth order whereas the subtracted one as perturbation. 
Equivalently, for calculating physical quantities this amounts to replace 
the bare propagators and vertices by resummed HTL Green 
functions~\cite{braaten}.
In this way, interesting quantities of the QGP have been 
computed~\cite{andersen,andersen0,andersen1,andersen1a,andersen2,andersen1b,braaten2,purn1,mgm} within the HTLpt. 
However, HTLpt suffers from the fact  
 that one uses HTL resummed Green functions
also for hard momenta of the order $T$. 
A systematic description of physical quantities requires an explicit 
separation of hard ($\sim T$) and
soft ($\sim gT$) scales~\cite{braaten}. Following this approach, 
a number of relevant physical 
quantities, such as signatures
of the QGP, has consistently been calculated~\cite{markus}.
A convincing reason for studying the QNS in HTL approximation 
is the fact
that this approach has also been used for the free energy of the QGP, 
where different implementations of the HTL method and their validity have 
been discussed. The QNS
may serve as another test for these approaches.

The usual definition of the QNS for a given flavor 
which measures the response of the 
quark number density $\rho_q$ to an infinitesimal change in the quark chemical 
potential $\mu_q +\delta \mu_q$ is given by~\cite{fluc,mcler,hatsuda}: 
\begin{equation}
\chi_q(T) = \left.\frac{\partial \rho_q}{\partial \mu_q}\right |_{\mu_q=0}.
\label{eq0}
\end{equation}
Since the quark (baryon) number is a conserved quantity, the QNS obeys a 
Kubo relation \cite{fluc} 
\begin{equation}
\chi_q(T) = \beta \int \ d^3x \ \left \langle j_0(0,{\vec x})j_0(0,{\vec 0})
\right \rangle , \label{eq1}
\end{equation} 
where $j_0$ is the time component of the quark current
$j_\mu=\bar \psi \gamma_\mu \psi$. The equivalence of the static susceptibility
(\ref{eq0}) and the static limit of the dynamical susceptibility
(\ref{eq1}) follows from \cite{fluc}
\begin{equation}
\rho_q=\frac{1}{V} 
\frac{{\rm{Tr}}\left [  {\cal N}_q e^{-\beta \left({\cal H}-\mu_q {\cal N}_q
\right )}\right ]}
{{\rm{Tr}}\left [e^{-\beta \left({\cal H}-\mu_q {\cal N}_q\right )}\right ]} =
\frac{\langle {\cal N}_q\rangle}{V} \ , \label{eq2}
\end{equation}
with the quark number operator ${\cal N}_q=\int d^3x j_0(0, {\vec x})$ and
volume $V$. The derivative with respect to $\mu_q$ in (\ref{eq2}) according to
(\ref{eq0}) agrees with (\ref{eq1}) {\it only if the number operator is a conserved 
quantity and hence commutes with the Hamiltonian} ${\cal H}$~\cite{fluc}. 

Introducing the vector meson correlator
$S_{\mu\nu}(t,{\vec x})= \langle j_\mu(t,{\vec x})j_\nu(0,{\vec 0})\rangle$
and taking the Fourier transform of 
$S_{00}(0,{\vec x})$, the QNS according to (\ref{eq1}) can be written
as~\cite{Das}
\begin{eqnarray}
\chi_q(T)  
&=& \ \ \beta \int d^3x \int \frac{d\omega}{2\pi} \int \frac{d^3p}{(2\pi)^3} 
e^{-i{\vec p}\cdot {\vec x}} \ S_{00}(\omega,p) \nonumber \\ 
&=& \beta \ \int _{-\infty }^{+\infty }\frac{d\omega }{2\pi }
\ \  S_{00}\left(\omega ,0\right) 
\, . \label{eq3}
\end{eqnarray}
This equation is known as a {\it thermodynamic sum rule} as it relates the
thermodynamic derivatives to the correlation function due to the 
symmetry of the system. Examples of such sum rules for various
systems can be found in Ref.~\cite{fluc}.
The fluctuation - dissipation theorem relates $S_{00}$ to the vector meson
self-energy $\Pi_{00}$ via \cite{fluc}
\begin{equation}
S_{00}(\omega,p)=\frac{2} {\exp(-\beta\omega)-1} 
{\rm{Im}} \Pi_{00}(\omega,p) \ \ . \label{eq2a}
\end{equation}
An example for this self-energy within the 1-loop order with
effective Green functions is shown in Fig.1, where the blobs indicate 
effective  quark propagators and quark-meson vertices.

The time independence of ${\cal N}_q$, corresponding to the continuity 
equation $\partial _\mu j^\mu=0$, implies that the quantity
$\int d^3x S_{00}(0,{\vec x})$  in (\ref{eq1}) is time independent.
This is guaranteed if $S_{00}\left(\omega ,0\right) \propto \delta(\omega)$ 
\cite{fluc}. 

\vspace*{-4cm}
\begin{figure}
\hspace*{-1cm}\epsfxsize=12cm
\epsfbox{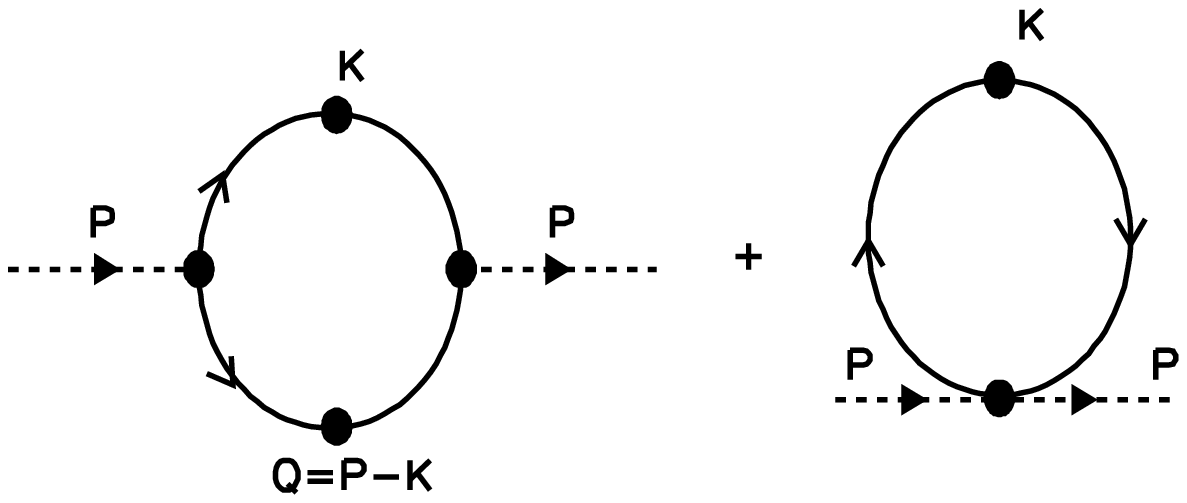}
\end{figure}

\vspace*{-5.5cm}

\noindent{Fig.1: The 1-loop vector meson self-energy diagrams.}

\vspace*{0.2cm}

In usual perturbation theory the equivalence of (\ref{eq0}) and (\ref{eq1})
follows from the fact that the loop expansion agrees with an
expansion in the coupling constant. For example, to lowest order the
number density follows from the tadpole diagram (containing only a
bare propagator and vertex) at finite temperature and
chemical potential, which corresponds
to the 1-loop polarization tensor of the vector correlator (\ref{eq2a}).
For the HTL resummation the loop expansion and the coupling expansion are
not equivalent as the first one includes higher order effects through 
resummation. This mixing can cause an incompleteness in 
physical quantities computed in a given loop order, as we shall 
discuss below. 

Now we would like to discuss 
how the QNS can be calculated consistently 
within the HTL approximation. Our starting point is the most 
general expression of the quark self-energy~\cite{weldon}, which enters 
the full quark propagator. In the rest frame of a medium in the 
chiral limit it reads
\begin{equation}
\label{eq5}
\Sigma \left( K\right) =-a(k_{0},k)K\! \! \! \! /-b(k_{0},k)\gamma ^{0}\, ,
\end{equation}
where the scalar quantities $a$ and $b$ 
are functions of the energy $k_0$ and of the magnitude 
of the spatial momentum, $k$, and $K=(k_0,{\vec k}\,)$. 
The full propagator is given by
\begin{equation}
S^{-1}_{F}\left( K\right) =K\! \! \! \! /-\Sigma \left( K\right) =
\left( 1+a\right) K\! \! \! \! /+b\gamma ^{0}\, . 
\end{equation}
Using the helicity representation \cite{braaten2} it can also be written as 
\begin{eqnarray}
S_{F}\left( k_{0},k\right)  & = & \frac{\gamma ^{0}-\widehat{k}\cdot
\overrightarrow{\gamma }}{2D_{+}\left( k_{0},k\right) }+\frac{\gamma ^{0}
+\widehat{k}\cdot \overrightarrow{\gamma }}{2D_{-}\left( k_{0},k
\right) }\nonumber\, ,\\
D_{\pm }\left( k_{0},k\right)  & = & (-k_{0}\pm k)(1+a)-b\, . \label{eq6}
\end{eqnarray}
Charge invariance demands $D_\pm(-k_0,k)=-D_\mp(k_0,k)$, which implies 
$a(-k_0,k)=a(k_0,k)$ and $b (-k_0,k)=-b(k_0,k)$.
The zeros of $D_\pm$ describe the in-medium 
propagation or {\it quasiparticle} (QP) dispersion relation 
of a particle excitation with energy $\omega_+(k)$ 
and of a mode called plasmino with energy, $\omega_-(k)$, which 
exists only in the medium \cite{braaten2,peshier}. 
In addition, $D_\pm$ contains a discontinuous part corresponding to
{\it Landau damping} (LD) 
if the quark self-energy has 
an imaginary part. 

Using these general properties of the quark propagator, the spectral 
representation of the in-medium propagator (\ref{eq6}) can be written as
\begin{eqnarray}
\rho_\pm(\omega,k) &=& -\frac{1}{\pi}{\rm{Im}} \frac{1}{D_\pm} = 
{\cal R}_\pm(\omega,k)\delta(
\omega-\omega_\pm) \nonumber \\
&& +{\cal R}_\mp(-\omega,k) \delta(\omega+\omega_\mp) +
\rho_\pm^{\rm{dis}}(\omega,k)
\ \ , \label{eq6p}
\end{eqnarray}
where ${\cal R}_\pm=({\rm d}D_\pm/{\rm d}\omega)^{-1}$ are the residues 
corresponding to the poles of the propagators, and 
 $\rho_\pm^{\rm{dis}}$ are discontinuities. In the HTL-approximation
the discontinuities, given explicitly e.g. in Ref.\cite{purn},
are located below the light cone ($k_0^2 < k^2$),
where Landau damping occurs. 
We also note that for each pole the spectral function will involve
a $\delta$-function.
The Ward identity relates the quark propagator to the quark-meson vertex
via
\begin{eqnarray}
(K_2-K_1)_\mu\Gamma^\mu(-K_2,K_1;K_2-K_1) && \nonumber \\
= S_F^{-1}(K_2)-S_F^{-1}(K_1).
\label{wi}
\end{eqnarray}
In the HTL approximation the effective vertices are discussed in 
Ref.\cite{purn}.

Using the effective propagators and vertices given in
(\ref{eq6}) and (\ref{wi}), the contribution of the first 
diagram in Fig.~1 can be written as
\begin{eqnarray}
{\rm Im}\, \Pi _{00}^s\left( \omega ,0\right)   &=&  4N_{f}
N_{c}\pi\left( 1-e^{\beta \omega }\right) \int \frac{d^{3}k}{\left
( 2\pi \right) ^{3}}\int dx\int dy  \nonumber \\
&\times& \delta \left( \omega -x-y\right) n_{F}\left( x\right)
n_{F}\left( y\right) \left \{ \left( 1-f(x,y)\right) ^{2}
 \right. \nonumber \\
&\times& \left. \rho _{+}( x,k\right) \rho _{-}\left( y,k\right) 
+ \frac{1}{\omega ^{2}\pi} \left [ {{\rm Im}\, F_+\left( x,k\right) } 
\right. 
\nonumber \\
&\times& \left. \left. \rho _{-}\left( y,k\right) 
+ {{\rm Im}\, F_-\left( y,k\right) } 
\rho _{+}\left( x,k\right) \right.  \right. \Big ] \Big \} ,  \label{eqn22}   
\end{eqnarray}
where $F_\pm(z,k)=D_\pm(z,k)+z\mp k$,  $z$ is the energy of the internal
quarks, $f(x,y)=(x+y)/\omega$, $N_f$ is the number of quark flavors, and 
$N_c$ is the number of colors.
The contribution of the tadpole diagram, 
${\rm Im}\, \Pi _{00}^t\left( \omega ,0\right)$, in Fig.~1 is identical apart 
from an 
opposite sign of the second term inside the curly braces in (\ref{eqn22}). 
Using the explicit form of the spectral functions (\ref{eq6p}), the
total contribution of Fig.~1 is
\begin{eqnarray}
&& {\rm Im}\, \Pi _{00}\left( \omega ,0\right)=
\left [{\rm Im}\, \Pi _{00}^s\left( \omega \right)+{\rm Im}\, 
\Pi _{00}^t\left( \omega \right)\right ] 
 =  4N_{f} N_{c}\pi \nonumber \\ && \left (1-e^{\beta \omega }\right) 
\delta(\omega) \int \frac{d^{3}k}{\left
( 2\pi \right) ^{3}} \left [ {\cal R}^{2}_{+}( \omega _{+},k) 
 n_{F} ( \omega _{+}) n_{F}( -\omega _{+}) \right. \nonumber \\ 
&&+ \left.    
{\cal R}^{2}_{-}( \omega _{-},k) n_{F} ( \omega _{-})
n_{F}(- \omega _{-}) \right ]  
 \ , \label{eqn23}
\end{eqnarray} 
where the LD contributions explicitly vanish. ${\rm Im}\Pi _{00}$  in 
(\ref{eqn23}) contains only 
the QP contributions following from the dispersion relations, $D_\pm=0$, and 
the requirement for the {\it conservation
law}, $S_{00}\propto \delta(\omega)$, is satisfied. This necessitates the
use of effective 3- and 4-point functions and  
also the modification of the charge operator, 
as $\bar \psi (\gamma_0 +\delta\Gamma_0)\psi$, by inclusion of
the effective 3-point function, in contrast to Ref.~\cite{rebhan1,rebhan}.
This non-local modification makes the charge operator different
from that of (\ref{eq0}) and (\ref{eq1}) but is essential to satisfy the
 Ward identity which is related to the gauge invariance and 
conservation law. 
This implies that the number density has to be calculated
with propagators and vertices related by Ward identities 
at finite $T$ and $\mu_q$. 

\vspace{-0.35in}
\begin{figure}
\epsfxsize=3.5in
\epsfbox{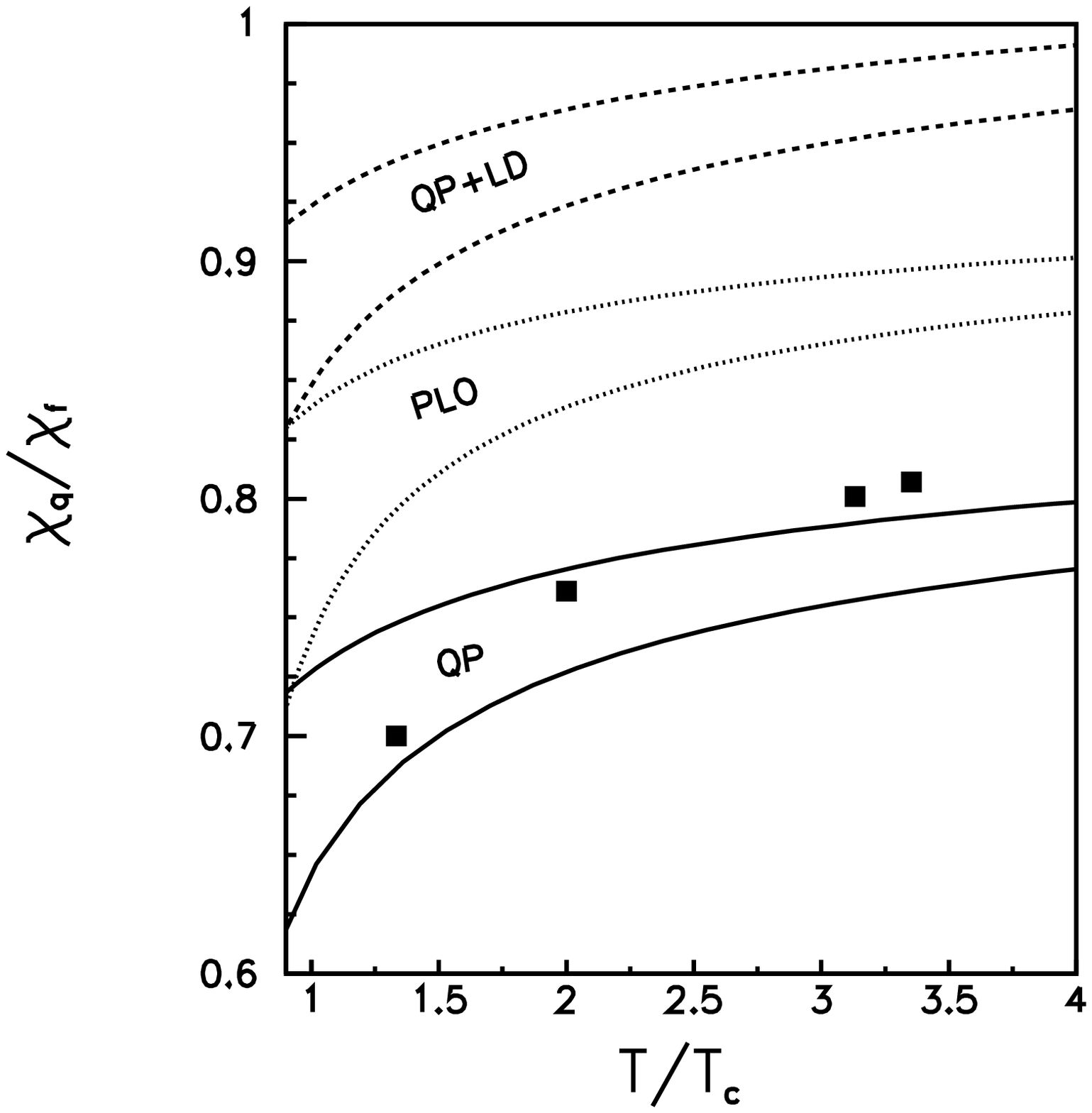}
\vspace{-1.0in}

\noindent{Fig.2: The QNS in different approximations (see text) divided by the
lowest order result (free QNS) as a function of $T/T_c$ with 
$T_c/\Lambda_{\overline{MS}}=0.49$ 
\cite{msb}. In each band the lower curve corresponds to the choice of
the renormalization scale 
${\bar \mu }=2\pi T$ 
and the upper one to ${\bar \mu} =4\pi T$. The 
squares represent recent lattice data~\cite{gavai1}.}
\end{figure}

Now the 1-loop QNS follows from combining (\ref{eqn23}) and
(\ref{eq3}) 
\begin{eqnarray}
\chi_q(T) &=& 4 N_f N_c \beta \int \frac{d^3k}{(2\pi)^3} \left [
{\cal R}_+^2(\omega_+,k) n_F(\omega_{+})
n_F(-\omega_{+})  \right. \nonumber \\
&+& \left. 
{\cal R}_-^2(\omega_-,k) n_F(\omega_{-}) n_F(-\omega_{-})
\right ] \, \,  , \label{eqn24}
\end{eqnarray}
which is identical with the one-loop HTLpt result~\cite{purn}, 
if the residues are 
obtained from the HTL resummed propagators. 
In this case the functions ${\cal R}_\pm$ in (\ref{eqn24}) are given by
\begin{equation}
{\cal R}_\pm(\omega_\pm ,k)=\frac{\omega_\pm^2(k)-k^2}{2m_q^2}
\label{eqn25}
\end{equation}
with the HTL quark dispersion relations $\omega_\pm(k)$ and the HTL 
effective quark mass $m_q=g(T)T/\sqrt{6}$.
The result (band between the solid line denoted by QP in Fig.~2)
agrees well with recent lattice data~\cite{gavai1}. (Here we used
the 2-loop result for the running coupling constant $g(T)$ in contrast to 
Ref.~\cite{purn}, where the 1-loop expression was adopted. For the 
renormalization scale $\bar \mu$ entering the running coupling constant,
we choose two different values leading to the bands in Fig.~2.)
However, 
it overcounts the leading order perturbative result (dotted band
denoted by PLO in Fig.~2) given as 
$\chi_q/\chi_f=(1-2\alpha_s/\pi+\cdots)$~\cite{rebhan}. 
We note that this can be 
cured by going to the calculation at 2-loop order. Recently, it has been shown
that the 2-loop order 
calculation~\cite{andersen2,andersen1b} for the 
thermodynamic potential reveals the correct inclusion of the leading order 
effects but happens to be very close to the 1-loop result.
 
The physical relevance of the plasmino branch has been discussed 
in connection with the dilepton production in relativistic heavy-ion 
collisions \cite{braaten2,peshier}. Using the HTL approach its
contribution to the QNS, i.e. the second term of (\ref{eqn24}), amounts only
to less than one percent of the total susceptibility. This is caused by the 
fact that within the HTL approximation the spectral strength of the plasmino
branch vanishes exponentially for hard momenta, $k\gg gT$, and that we 
integrate in (\ref{eqn24}) over all momenta.
 
Now, employing only resummed propagators and bare vertices as in 
Ref.~\cite{rebhan1}, the QNS
is obtained from (\ref{eqn22}) by setting the vertex correction functions,
$f(x,y)$ and $F_\pm(x,y)$, equal to zero. Then the QNS contains LD 
contributions coming from
the discontinuity of the 2-point HTL function in addition to the QP 
contribution in (\ref{eqn24}). The result is shown in Fig.~2 by the dashed band
denoted by QP+LD which clearly overestimates the lattice data. 
Although the charge operator is local in this approach~\cite{rebhan}, 
$S_{00}(\omega,0)$ is not proportional to $\delta(\omega)$, implying  
that $d\rho_q/dt\neq 0$.
This amounts to a violation of the Ward identity,
{\it i.e}, gauge invariance, for the effective HTL Green functions and 
also sacrifices the conservation law and hence the thermodynamic sum rule.
Note, however, that it has been argued that the gauge dependence of the
2-loop effective action within the $\Phi$-derivable approximation affects only
higher order contributions~\cite{arri}.

In summary, we have discussed that the QNS can be 
computed consistently within the HTLpt 
in order to satisfy the requirement of the thermodynamic 
sum rule. This requires to consider 3-point 
and 4-point HTL functions, related to the HTL propagator by Ward identities.
As discussed above, the QNS within the one-loop order overestimates the PLO, 
which,
however, can be cured by extending the calculation to 2-loop orders in HTLpt.
This would, indeed, be a challenging as well as nontrivial task due to its 
complexity. However, if it turns out to be a small correction,{\it e.g.}, 
like in the case of the 
quark-gluon thermodynamic potential, the 1-loop HTLpt 
result presented here, agreeing well with lattice data,
would be a useful approximation.

\vspace{0.15in}

\noindent{\it Acknowledgment:} M.G.M. is thankful to  C. Greiner,
S. Juchem, S. Leupold and A. Peshier for various illuminating discussions
and to the AvH Foundation and GSI (Dramstadt) for financial support. M.H.T.
was supported by DLR (BMBF) under grant no. 50WM0038.

\end{document}